\documentclass[%showpacs,
showkeys,12pt,
preprint,preprintnumbers,nofootinbib,
groupedaddress,superscriptaddress,amsmath,amssymb]{revtex4}
\usepackage{graphicx}% Include figure files
\usepackage{dcolumn}% Align table columns on decimal point
\usepackage{bm}% bold math
\usepackage{amssymb}
\usepackage{amsmath}
\usepackage{epsfig}    
\usepackage{color}
\usepackage{slashed}
\usepackage{hhline}
\def\be{\begin{equation}}
\def\ee{\end{equation}}
\newcommand{\bea}{\begin{eqnarray}}
\newcommand{\eea}{\end{eqnarray}}
\newcommand{\nn}{\nonumber}

\numberwithin{equation}{section}

\begin{document}

{\begin{flushright}{KIAS-P16011}
\end{flushright}}

%%%%%%%%%
\title{Generalized Zee-Babu model with 750 GeV Diphoton Resonance}
%\preprint{KIAS-P14078}
%
\author{Takaaki Nomura}
\email{nomura@kias.re.kr}
\affiliation{School of Physics, KIAS, Seoul 130-722, Korea}

\author{Hiroshi Okada}
\email{macokada3hiroshi@gmail.com}
\affiliation{Physics Division, National Center for Theoretical Sciences, Hsinchu, Taiwan 300}

\date{\today}

\begin{abstract}
We propose a generalized Zee-Babu model with a global $U(1)$ B-L symmetry, in which we classify the model in terms of the number of the hypercharge $N/2$ of the isospin doublet exotic charged fermions. Corresponding to each of the number of $N$, we need to introduce some multiply charged bosons in order to make the exotic fields decay into the standard model fields.   
We  also discuss the muon anomalous magnetic moment and the diphoton excess depending on $N$, and we show what kind of models are in favor of these phenomenologies.
\end{abstract}
\maketitle
\newpage

\section{Introduction}
{
The recent measurements reported by ATLAS and CMS experiments implies that a new particle ($\Phi_{\rm New}$) might exist at around 750 GeV by the diphoton invariant mass spectrum from the run-II data in 13 TeV~\cite{ATLAS-CONF-2015-081,CMS:2015dxe}.
And a typical interpretation is known as
%in terms as   This can typically be interpreted as the following 13 TeV data in terms of the production cross section of $\Phi_{\rm New}$ and its branching ratio of two photons,
\begin{align}
&\mu_{\rm ATLAS}=\sigma(2p\to \Phi_{\rm New})\times {BR}(\Phi_{\rm New} \to 2\gamma)=(6.2^{+2.4}_{-2.0})\ {\rm fb},\\
&\mu_{\rm CMS}=\sigma(2p\to \Phi_{\rm New})\times {BR}(\Phi_{\rm New} \to 2\gamma)=(5.6\pm{2.4})\ {\rm fb}.
\label{lhc-exp}
\end{align}
Furthermore the ATLAS experiment group~\cite{ATLAS-CONF-2015-081} announced $\Gamma_{\Phi_{\rm New}}=45$ GeV, which is the best fit value of the total decay width of $\Phi_{\rm New}$, while the CMS experiment group~\cite{CMS:2015dxe} indicated a rather smaller decay width.
%, and $\Gamma_{\Phi_{\rm New}}=5.3$ GeV is given as the experimental resolution obtained by the analysis~\cite{McDermott:2015sck}.
 It might suggest that $\Phi_{\rm New}$ be a scalar (or pseudoscalar) and additional new fields with nonzero electric charges are in favor of being introduced, since sizable branching fraction for $\Phi_{\rm New} \to \gamma \gamma$ requires $\Phi_{\rm New}$ strongly interacts with such charged fields.
In order to provide reasonable explanations (or interpretations),
}
a vast of paper along this line of issue has been recently arisen in Ref.~\cite{Harigaya:2015ezk, Mambrini:2015wyu,Backovic:2015fnp,Angelescu:2015uiz,Nakai:2015ptz,Knapen:2015dap,Buttazzo:2015txu,Pilaftsis:2015ycr,Franceschini:2015kwy,DiChiara:2015vdm,Higaki:2015jag,McDermott:2015sck,Ellis:2015oso,Low:2015qep,Bellazzini:2015nxw,Gupta:2015zzs,Petersson:2015mkr,Molinaro:2015cwg,Dutta:2015wqh,Cao:2015pto,Matsuzaki:2015che,Kobakhidze:2015ldh,Martinez:2015kmn,Cox:2015ckc,Becirevic:2015fmu,No:2015bsn,Demidov:2015zqn,Chao:2015ttq,Fichet:2015vvy,Curtin:2015jcv,Bian:2015kjt,Chakrabortty:2015hff,Ahmed:2015uqt,Agrawal:2015dbf,Csaki:2015vek,Falkowski:2015swt,Aloni:2015mxa,Bai:2015nbs,Gabrielli:2015dhk,Benbrik:2015fyz,Kim:2015ron,Alves:2015jgx,Megias:2015ory,Carpenter:2015ucu,Bernon:2015abk,Chao:2015nsm,Arun:2015ubr,Han:2015cty,Chang:2015bzc,Chakraborty:2015jvs,Ding:2015rxx,Han:2015dlp,Han:2015qqj,Luo:2015yio,Chang:2015sdy,Bardhan:2015hcr,Feng:2015wil,Antipin:2015kgh,Wang:2015kuj,Cao:2015twy,Huang:2015evq,Liao:2015tow,Heckman:2015kqk,Dhuria:2015ufo,Bi:2015uqd,Kim:2015ksf,Berthier:2015vbb,Cho:2015nxy,Cline:2015msi,Bauer:2015boy,Chala:2015cev,Barducci:2015gtd,Boucenna:2015pav,Murphy:2015kag,Hernandez:2015ywg,Dey:2015bur,Pelaggi:2015knk,deBlas:2015hlv,Belyaev:2015hgo,Dev:2015isx,Huang:2015rkj,Moretti:2015pbj,Patel:2015ulo,Badziak:2015zez,Chakraborty:2015gyj,Cao:2015xjz,Altmannshofer:2015xfo,Cvetic:2015vit,Gu:2015lxj,Allanach:2015ixl,Davoudiasl:2015cuo,Craig:2015lra,Das:2015enc,Cheung:2015cug,Liu:2015yec,Zhang:2015uuo,Casas:2015blx,Hall:2015xds,Han:2015yjk,Park:2015ysf,Salvio:2015jgu,Chway:2015lzg,Li:2015jwd,Son:2015vfl,Tang:2015eko,An:2015cgp,Cao:2015apa,Wang:2015omi,Cai:2015hzc,Cao:2015scs,Kim:2015xyn,Gao:2015igz,Chao:2015nac,Bi:2015lcf,Goertz:2015nkp,Anchordoqui:2015jxc,Dev:2015vjd,Bizot:2015qqo,Ibanez:2015uok,Chiang:2015tqz,Kang:2015roj,Hamada:2015skp,Huang:2015svl,Kanemura:2015bli,Kanemura:2015vcb,Low:2015qho,Hernandez:2015hrt,Jiang:2015oms,Kaneta:2015qpf,Marzola:2015xbh,Ma:2015xmf,Dasgupta:2015pbr,Jung:2015etr,Potter:2016psi,Palti:2016kew,Nomura:2016fzs,Han:2016bus,Ko:2016lai,Ghorbani:2016jdq,Palle:2015vch,Danielsson:2016nyy,Chao:2016mtn,Csaki:2016raa,Karozas:2016hcp,Hernandez:2016rbi,Modak:2016ung,Dutta:2016jqn,Deppisch:2016scs,Ito:2016zkz,Zhang:2016pip,Berlin:2016hqw,Bhattacharya:2016lyg,D'Eramo:2016mgv, Sahin:2016lda,Fichet:2016pvq, Borah:2016uoi,Stolarski:2016dpa,Fabbrichesi:2016alj,Hati:2016thk,Ko:2016wce,Cao:2016udb,Yu:2016lof,Ding:2016ldt,Alexander:2016uli,Davis:2016hlw,Dorsner:2016ypw,Faraggi:2016xnm,Djouadi:2016eyy,Ghoshal:2016jyj,Nomura:2016seu,Chao:2016aer,Han:2016bvl,Okada:2016rav,Franzosi:2016wtl,Martini:2016ahj,Cao:2016cok,Chiang:2016ydx,Aydemir:2016qqj,Abel:2016pyc}.

%\textcolor{blue}{
Zee-Babu type \cite{zee-babu} of 
radiative seesaw models could provide one of the economical scenarios to include such new exotic fields with nonzero electric charges (bosons or fermions) that are naturally introduced in order not only to explain the diphoton excess but also  to explain the tiny neutrino masses. Also the model can easily be extended to the multi-charged bosons and fermons.
% Moreover there often exist dark matter (DM) candidates, which also  play a role in generating tiny neutrino masses.}

In our paper, we propose a generalized Zee-Babu model with a global $U(1)$ B-L symmetry, in which neutrino masses are induced at the two loop level. Here we introduce vector like isospin doublet fermions with general $N/2$ hypercharges that can explain the discrepancy of the muon anomalous magnetic moment to the standard model (SM) sizably.  Corresponding to each of the number of $N$, we need to introduce some multiply charged bosons in order to make the exotic fields decay into the standard model fields.
Diphoton excess is explained by introducing several charged bosons, depending on the number of $N$.
Here we classify the model according to the number of $N$, and we discuss what kind of models  are in favor of explaining the sizable diphoton resonance as well as the muon anomalous magnetic moment. 
%Also we discuss the decay processes of the new exotic fields into the SM fields, depending on the number of $N$,

%

%, (slightly depending on the loop function).
%, which is likely to type-II model at the tree level~\cite{type-2}.
%This paper is organized as follows.
%\textcolor{red}{
In Sec.~II, we show our model,  including neutrino sector, and muon anomalous magnetic moment.
In Sec.~III, we discuss the decay processes of  exotic fields.
In Sec.~IV, we discuss the diphoton excess. We conclude and discuss in Sec.~V.
%}
%In appendices, we show the explicit Higgs potential and ...
%\newpage

%%%%%%%%%%%%%%%%%%%%%%%%%%%%%%%%%%%%%
%\section{The Model}
%\subsection{Model setup}
\section{ Model setup and Analysis}
 \begin{widetext}
\begin{center} 
\begin{table}[tbc]
%\begin{tiny}
\begin{tabular}{|c||c|c|c|c||c|c|c|c|}\hline\hline  
&\multicolumn{4}{c||}{Lepton Fields} & \multicolumn{4}{c|}{Scalar Fields} \\\hline
& ~$L_L$~ & ~$e_R^{}$~ & ~$L'^{}_{}$ ~ & ~$E$ ~  & ~$\Phi_{}$~  & ~$h^{+M}$  & ~$k^{+2M}$  & ~$\varphi$ \\\hline 
$SU(2)_L$ & $\bm{2}$  & $\bm{1}$  & $\bm{2}$  & $\bm{1}$ & $\bm{2}$ & $\bm{1}$   & $\bm{1}$    & $\bm{1}$ \\\hline 
$U(1)_Y$ & $-1/2$ & $-1$  & $-N/2$ & $-M$& $1/2$ & $+M$ & $+2M$  & $0$  \\\hline
$U(1)_{\rm B-L}$ & $-1$ & $-1$  & $-3$ & ${-3}$ & ${0}$ & ${2}$ & $6$  & $-2$  \\\hline
% $Z_2$ & $+$ & $+$  & $+$ & ${+}$  & $+$ & ${+}$ & $+$  & $-$ & $+$  & $+$  \\\hline
\end{tabular}
\caption{Contents of fermion and scalar fields
and their charge assignments under $SU(2)_L\times U(1)_Y\times U(1)_{\rm B-L}$, where $M$ is defined by $M=(1+N)/2$ with $N(=1,3,5,...)$ is the odd number.}
\label{tab:1}
% \end{tiny}
\end{table}
\end{center}
\end{widetext}
%
%We discuss a two-loop induced radiative neutrino model. 
In this section, we explain our model with a global  $U(1)_{\rm B-L}$ symmetry. 
The particle contents and their charges are shown in Tab.~\ref{tab:1}. As for the fermions,
we add some vector-like exotic isospin doublet charged fermions $L'$ with $-N/2$ hypercharge and isospin singlet fermions $E'$ with $-N/2$ hypercharge, where $N(=1,3,5,...)$ is generally an arbitrary odd number. As for the scalars, we introduce a $\pm M(\neq 0)$ electric charged scalar $h^{ \pm M}$, and a $\pm 2M$ charged scalar  $k^{\pm 2M}$ with different $U(1)_{B-L}$ quantum numbers, and a neutral scalar $\varphi$ in addition to the SM.
%%%
 Notice here that the electric charge to the fields with the $-N/2$ hypercharge, {\it i.e.}, $L'$, is given as 
\begin{align}
Q_{L'}=\left[\frac12,-\frac12\right]^T + \left[-\frac N2,-\frac N2\right]^T=\left[\frac{1-N}{2}, -\frac{1+N}{2}\right]^T\equiv [-M,-M-1]^T.
\end{align}
Thus we define $L'$ as 
\begin{align}
L'\equiv [\psi^{-M}, \psi^{-M-1}]^T.
\end{align}

We assume that only the Higgs doublet $\Phi$  and $\varphi$ have vacuum expectation values (VEVs), which are respectively symbolized by $v/\sqrt2$ and $v'/\sqrt2$.

%{\color{blue}
The relevant Lagrangian and Higgs potential under these symmetries are given by
\begin{align}
-\mathcal{L}_{Y}
&\supset
y_{\ell} \bar L_{L} \Phi e_{R} +f_{} \bar L_{L} L'_{R} h^{+M} +g_{L/R} \bar E^c_{L/R} E_{L/R}  k^{+2M}  
+h_{L/R}\bar L'_{L/R} \tilde\Phi E_{R/L} 
\nn\\
&+ { M_{L}} \bar L'_{L} L'_{R}  + { M_{E}} \bar E_{L} E_{R} +{\rm c.c.}, \label{Eq:lag-flavor}   \\
V =& \ m_\varphi^2 |\varphi|^2 + m_{h}^2 |h^{+M}|^2 + m_{k}^2 |k^{+2M}|^2 + m_{\Phi}^2 |\Phi|^2 
+\lambda_0 (k^{+2M} h^{-M} h^{-M} \varphi + {\rm c.c.})
\nonumber \\
& + \lambda_{\varphi}|\varphi|^4  + \lambda_{h}|h^{+M}|^4 + \lambda_{k}|k^{+2M}|^4 + \lambda_{\Phi}|\Phi|^4\nn\\
&
+\lambda_{\varphi h}|\varphi|^2 |h^{+M}|^2
+\lambda_{\varphi k}|\varphi|^2 |k^{+2M}|^2
+\lambda_{\varphi \Phi}|\varphi|^2 |\Phi|^2
+\lambda_{h k}|h^{+M}|^2|k^{+2M}|^2\nn\\
&
+\lambda_{h \Phi}|h^{+M}|^2 |\Phi|^2
+\lambda_{k\Phi}|k^{+2M}|^2 |\Phi|^2
%%%
 \label{eq:potential}
, 
\end{align}
where $\tilde\Phi\equiv i\sigma_2 \Phi^*$, the flavor indices are abbreviate and all the coulings are assume to be positive and real for brevity.
After the global $U(1)_{\rm B-L}$ spontaneous breaking by $\langle \varphi \rangle =v'/\sqrt{2}$, we obtain the trilinear term $\frac{\lambda_0 v'}{\sqrt2} k^{+2M} h^{-M} h^{-M}$ that plays an role in generating the neutrino masses at the two loop level.
%%%
The first term of $\mathcal{L}_{Y}$ generates the SM
charged-lepton masses $m_\ell\equiv y_\ell v_1/\sqrt2$ after the spontaneous breaking of electroweak symmetry by $\langle \Phi \rangle = v/\sqrt{2}$.
The isospin doublet scalar field is parameterized as $\Phi=[w^+,\frac{v+h_1 + iz}{\sqrt2}]^T$
where $v~\simeq 246$ GeV is VEV of the Higgs doublet, and the component of $w^\pm$
and $z$ are respectively absorbed by the longitudinal component of $W$ and $Z$ boson.
The isospin singlet scalar field can be parameterized by $\varphi=\frac{v'+h_2}{\sqrt2}e^{-2iG/v'}$.
Then we consider  the mixing of the CP-even Higgses, where the mass eigenstates are given by
\begin{equation}
\begin{pmatrix} h_1 \\ h_2 \end{pmatrix} = \begin{pmatrix} \cos \alpha & - \sin \alpha \\ \sin \alpha & \cos \alpha \end{pmatrix} \begin{pmatrix} H \\ h \end{pmatrix}.
\end{equation}
Here  $h$ and $H$ denote SM Higgs and heavier CP-even Higgs respectively.

\if0
In our analysis of $H$ production via gluon fusion, we focus on the Yukawa interactions of $H$ and top quark  
\begin{align}
{\cal L}^Y \supset &  -   \frac{m_{t}}{v} \frac{\sin \alpha}{\sin \beta} \bar t t H,
\end{align}
where $\tan \beta = v_2/v_1$ as usual.
The $h W^+W^-/ZZ$ and $H W^+W^-/ZZ$ couplings are respectively proportional to $\sin (\alpha - \beta)$ and  $\cos (\alpha - \beta)$ in the 2HDM~\cite{Gunion:1989we}. 
In this paper, we assume alignment limit~\cite{Carena:2013ooa}, $\alpha - \beta = \pi/2$, to suppress $H \to W^+W^-/ZZ$ decay channel.
\fi

{\it Exotic Charged Fermion mass matrix}:
%%%
The exotic charged fermion mass matrix with ${\pm M}$ electric charges is given by 
\begin{align}
-{\cal L}_{\text{mass}} &= (\overline{E^{-M}},\overline{\psi^{-M}})
\begin{pmatrix}
M_E &   m'_{} \\
m'_{} & M_L
\end{pmatrix}
\begin{pmatrix}
E^{-M} \\
\psi^{-M}
\end{pmatrix} + \text{h.c.}=
(\overline{E_1^{-M}},\overline{E_2^{-M}})
\begin{pmatrix}
M_{E^1} &   0 \\
0  & M_{E^2}
\end{pmatrix}
\begin{pmatrix}
E_{1}^{-M} \\
E_{2}^{-M}
\end{pmatrix} + \text{h.c.} , 
\end{align}
where we assume to be $m'_{} =  \frac{v}{\sqrt{2}} h_L =  \frac{v}{\sqrt{2}} h_R$ and $h^T=h=h_L=h_R$ for simplicity. 
The mass eigenstates $E^1$ and $E^2$ are defined by the bi-unitary transformation:
\begin{align}
\begin{pmatrix}
E^{-M} \\
\psi^{-M}
\end{pmatrix}
=
\begin{pmatrix}
c_{\theta_E} & -s_{\theta_E} \\
s_{\theta_E} &c_{\theta_E}
\end{pmatrix}
\begin{pmatrix}
E_1^{-M} \\
E_2^{-M}
\end{pmatrix}, 
\end{align}
where $s_{\theta_E} \equiv \sin\theta_E$ and $c_{\theta_E} \equiv \cos\theta_E$.
The mass eigenvalues and the mixing angles $\theta_E$ are respectively given by 
\begin{align}
M_{E^{1,2}} &= \frac{1}{2}\left(M_E + M_L \mp\sqrt{(M_E - M_L)^2 + 4m'^2}\right), \quad
\tan2\theta  = \frac{2 m'}{M_{E} - M_L},
\end{align}
where we define $M_{E^{1}} < M_{E^{2}}$, and the mass of the $\pm (M+1)$ electric charged fermion $\psi^{\pm(M+1)}$ is given by $M_L$.

%\subsection{Neutrino mass matrix}
%%%%%%%%%%%%%%%%%%%
\begin{figure}[tb]
\begin{center}
\includegraphics[width=100mm]{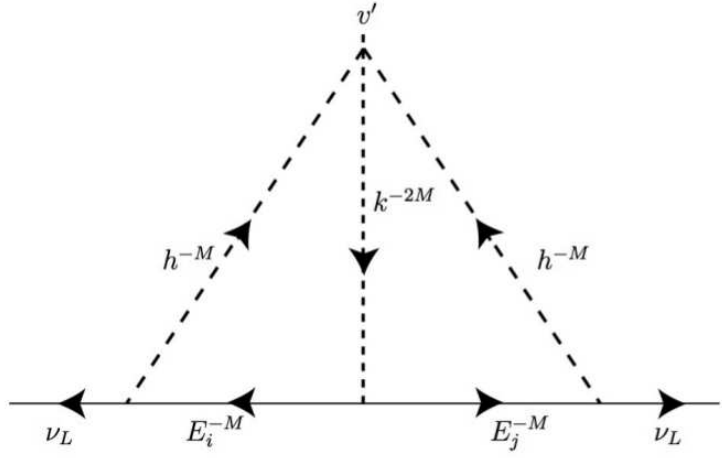}
\caption{ Neutrino masses at the two-loop level, where each $i,j$ runs 1 to 2. 
%%%
}   \label{fig:neut1}
\end{center}\end{figure}
%%%%%%%%%%%%%%%%%%%
{\it Neutrino mass matrix}:

%%%
The leading contribution to the active neutrino masses $m_\nu$  is given at two-loop level as shown in Figure~\ref{fig:neut1}, and its formula~\cite{Okada:2014qsa} is given as follows:
%\begin{widetext}
\begin{align}
(m_{\nu})_{ij}&= -\frac{\sqrt2 \lambda_0 v'(\sin2\theta)^2 f_{i\alpha} g^*_{\alpha\beta} f_{j\beta} }{4(4\pi)^4} \Pi_2,\\
\Pi_2&\equiv 
\int dxdydz\frac{\delta(a+y+z-1)}{z^2-z}\int d\alpha d\beta d\gamma \delta(\alpha+\beta+\gamma-1)\nn\\
&\times
\left[
z\ln\left[\frac{\Delta(E_{1\alpha},E_{2\beta})}{\Delta(E_{1\alpha},E_{1\beta})}\right]
+z\ln\left[\frac{\Delta(E_{2\alpha},E_{1\beta})}{\Delta(E_{2\alpha},E_{2\beta})}\right]\right. \\
&\left.
+M_{E_{1\alpha}}\left[\frac{M_{E_{1\beta}}}{\Delta(E_{1\alpha},E_{1\beta})} 
-
\frac{M_{E_{2\beta}}}{\Delta(E_{1\alpha},E_{2\beta})}\right]
%%%
+M_{E_{2\alpha}}\left[\frac{M_{E_{2\beta}}}{\Delta(E_{2\alpha},E_{2\beta})} 
-
\frac{M_{E_{1\beta}}}{\Delta(E_{2\alpha},E_{1\beta})}\right]
\right],\nn\\
%%%
\Delta(x,y)&\equiv
-\alpha\frac{x M^2_{x} + y m^2_{h^{\pm M}} +z m^2_{k^{\pm2M}} }{z^2-z} +\beta M^2_{y} + \gamma m^2_{h^{\pm M}},
\label{mnu1}
\end{align}%\end{widetext}
where %we define $M_{\rm max}={\rm Max}[M_{E^2}, m_{h^{\pm M}},m_{k^{\pm2M}}]$.
$m_\nu$ should be $0.001\ {\rm eV}\lesssim m_\nu \lesssim 0.1\ {\rm eV}$ from the neutrino oscillation data~\cite{pdf}.
Reminding the original Zee-Babu model, the loop function $\Pi_2$ is the order 1. Once we fix to be $\Pi_2 =1$, 
we obtain the following parameter region to satisfy the neutrino mass scale as
\begin{align}
71\ {\rm eV}\lesssim \lambda_0 v'(\sin2\theta)^2 f_{i\alpha} g^*_{\alpha\beta} f_{j\beta}\lesssim 7.1\times 10^3\ {\rm eV},
\end{align}
 which can easily be realized due to a lot of free parameters.

%\subsection{Muon anomalous magnetic moment}
{\it Muon anomalous magnetic moment}:
\begin{figure}[tb]
\begin{center}
\includegraphics[width=50mm]{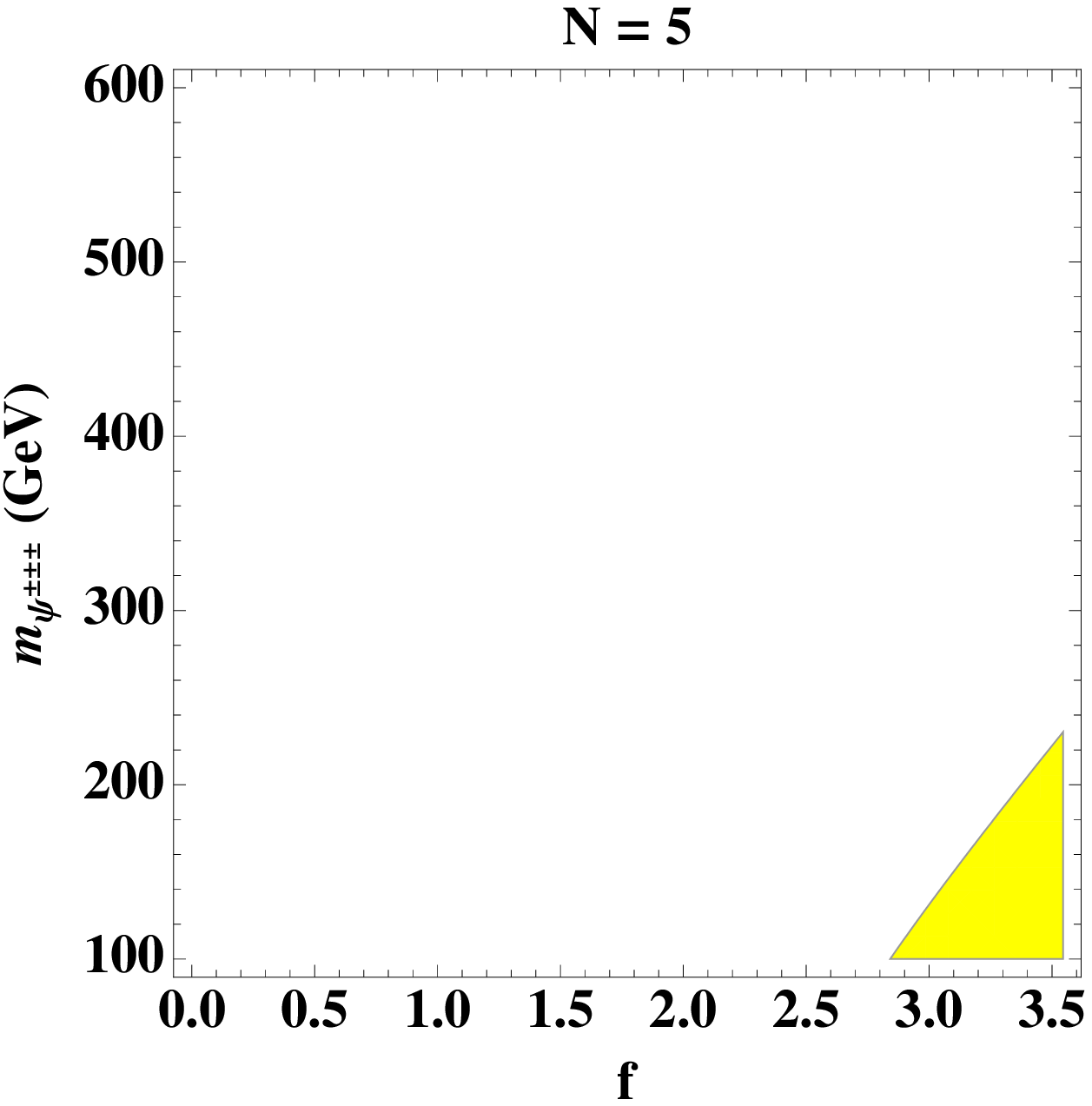}
\includegraphics[width=50mm]{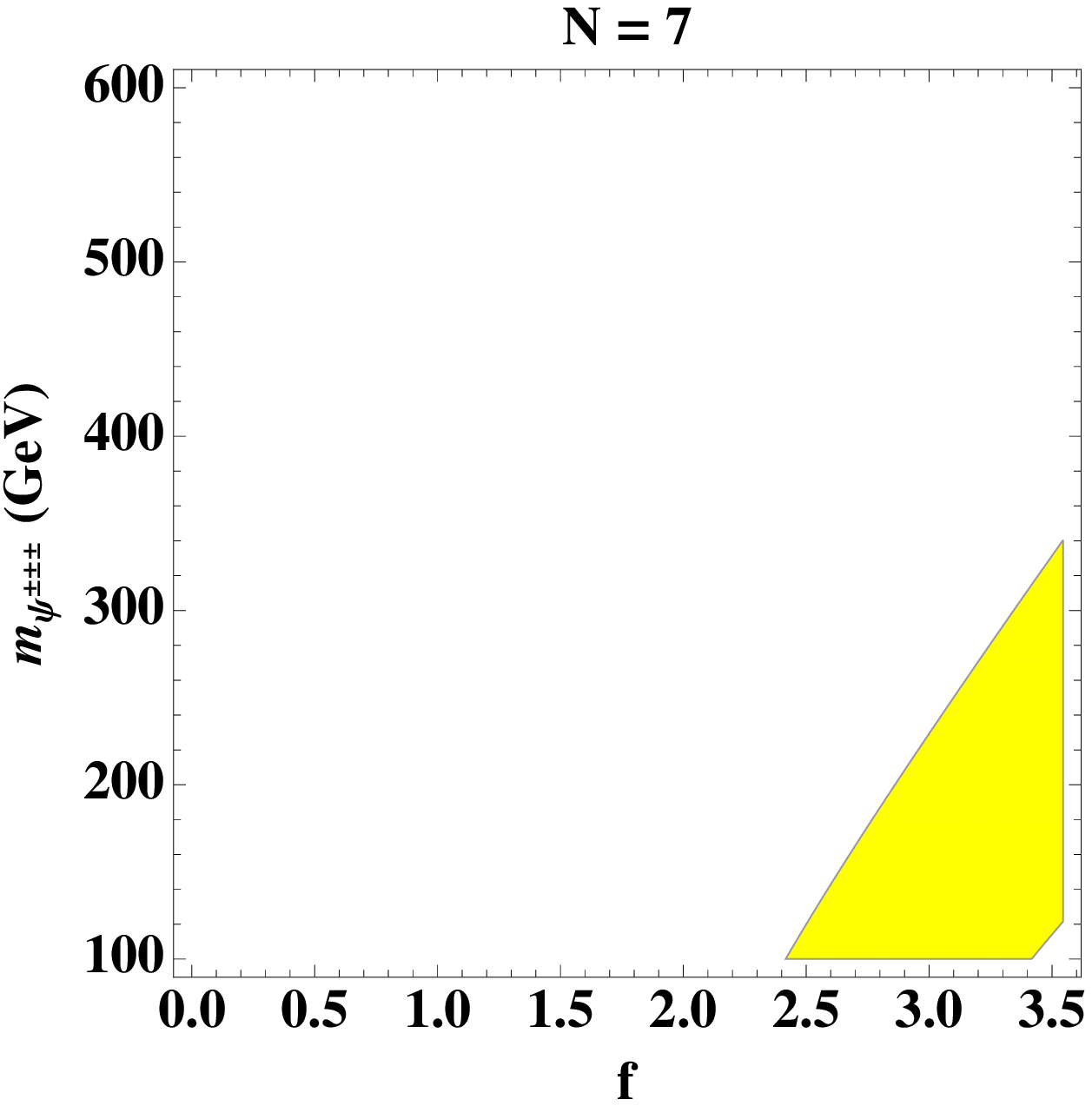}
\includegraphics[width=50mm]{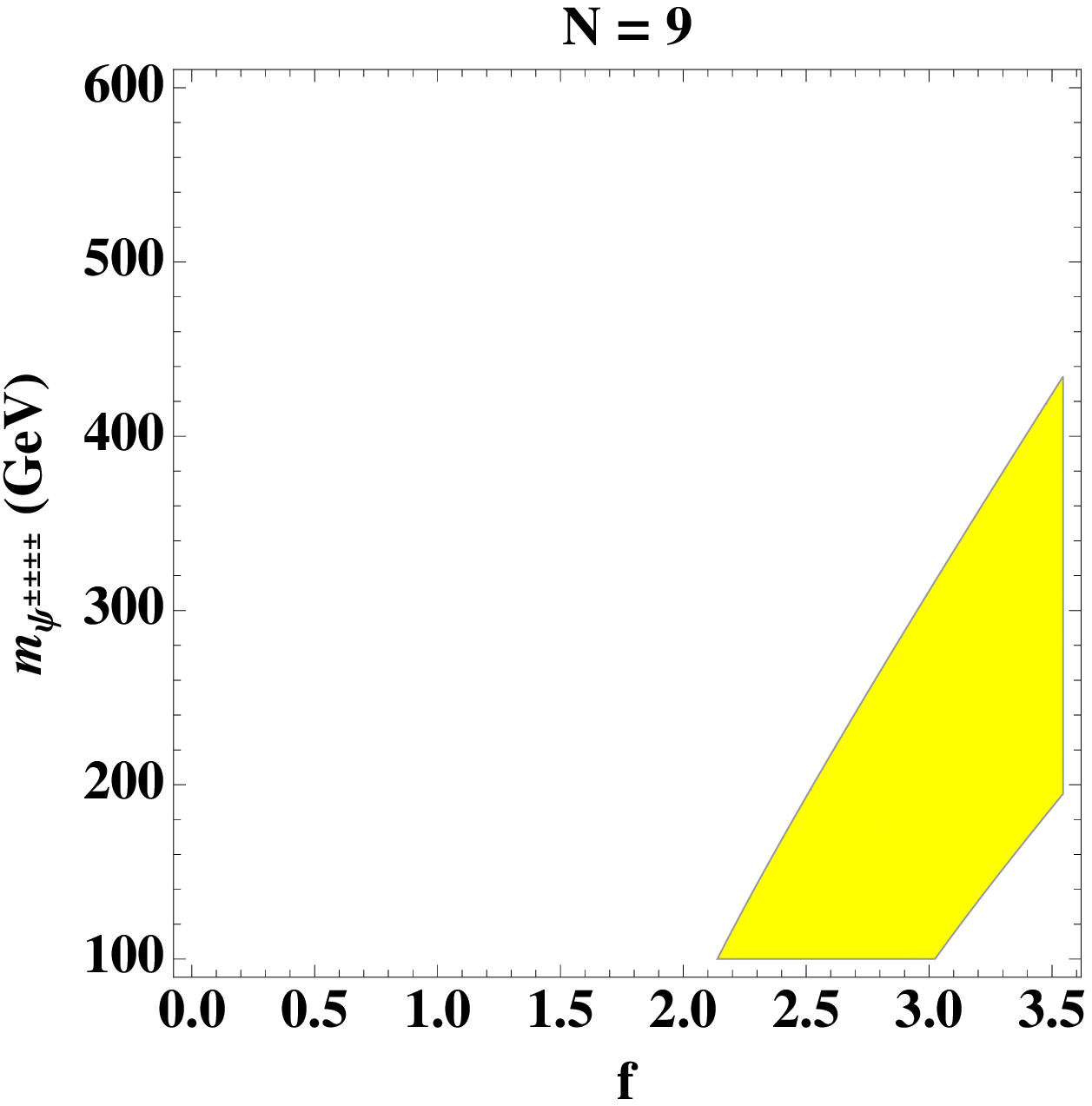}
\includegraphics[width=50mm]{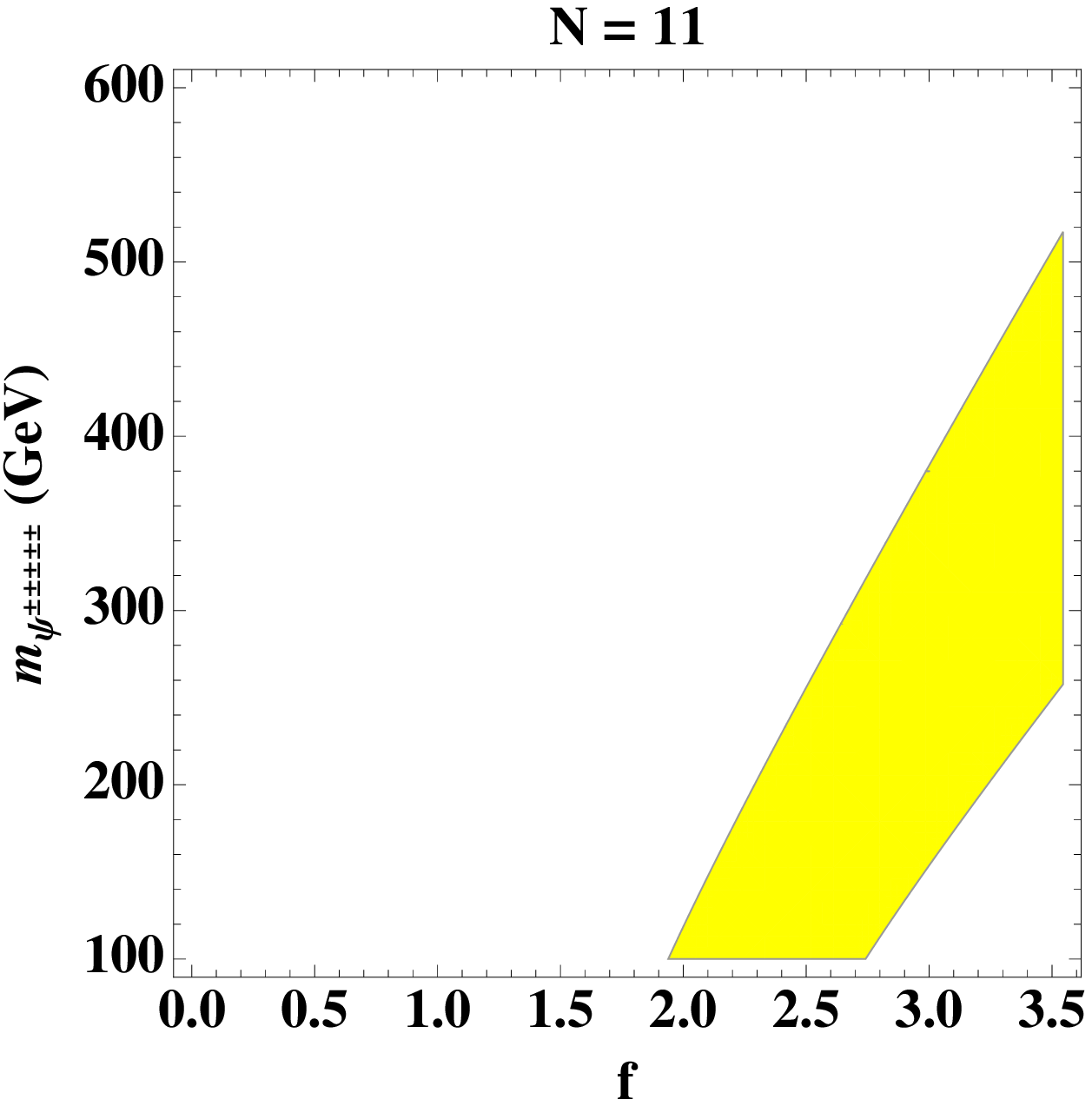}
\includegraphics[width=50mm]{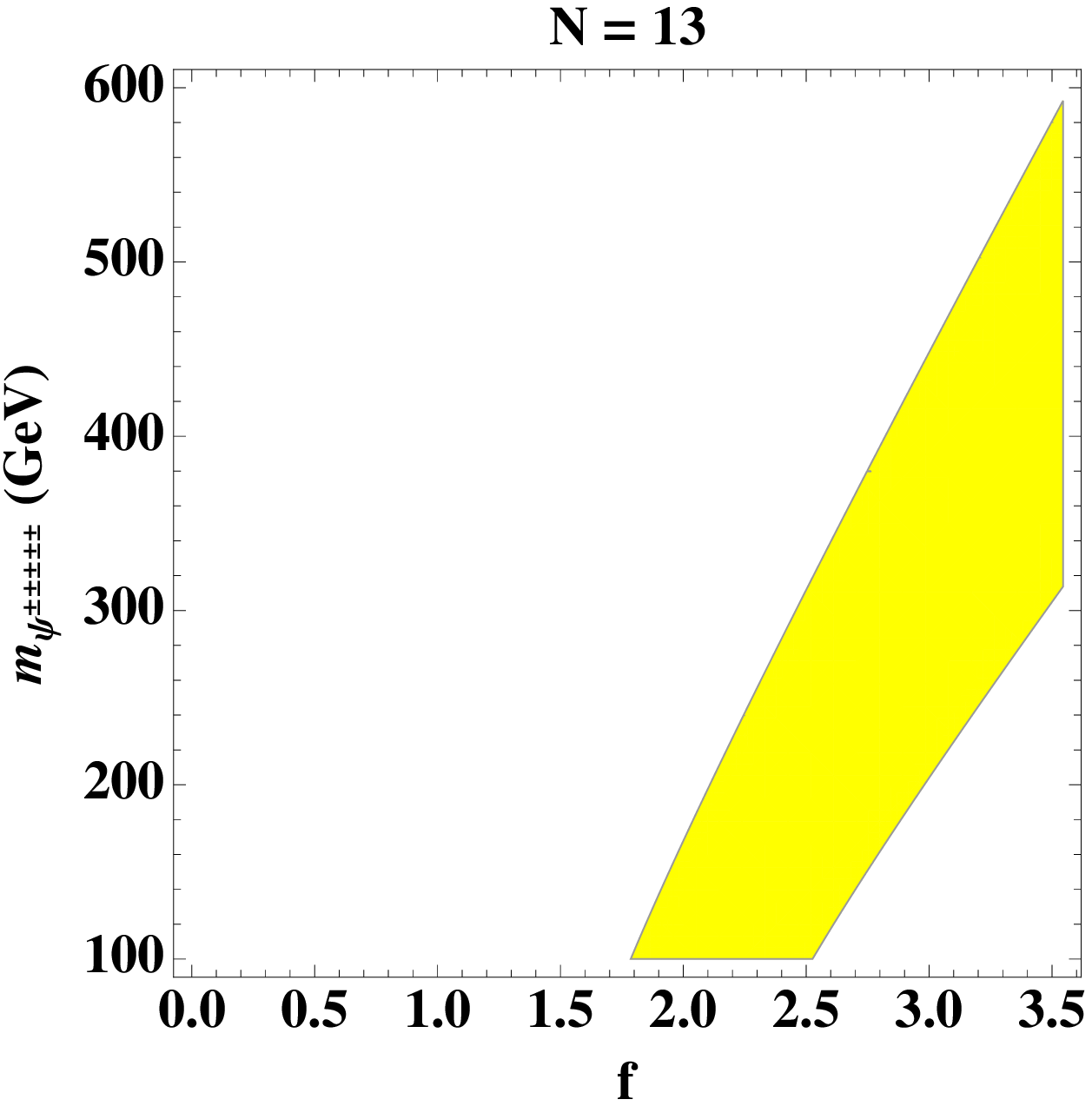}
\caption{The region plot in terms of $f$ and $M_{\psi^{\pm(M+1)}}$ plane for $N=5,7,9,11,13$ cases, where we fix $m_{h^{\pm M}}=380$ GeV and $N=3$ does not have the allowed region. 
The yellow region satisfies the sizable muon anomalous magnetic moment $2.0\times 10^{-9} \lesssim \Delta a_\mu\lesssim 4.0\times 10^{-9}$.}
\label{fig:muong-2}
\end{center}\end{figure}
%%%
%The muon anomalous magnetic moment (muon $g-2$) has been measured at Brookhaven National Laboratory
%The current average of the experimental results is given by~\cite{bennett}
%\begin{align}a^{\rm exp}_{\mu}=11 659 208.0(6.3)\times 10^{-10}. \notag\end{align}
%It has been well known 

{Brookhaven National Laboratory has announced 
a discrepancy between the experimental data and the prediction in the SM, and 
%that suggests there is a discrepancy between the experimental data and the prediction in the SM. 
its difference, which is denoted by $\Delta a_{\mu}\equiv a^{\rm exp}_{\mu}-a^{\rm SM}_{\mu}$,
is respectively given
}
 in Ref.~\cite{discrepancy1} and Ref.~\cite{discrepancy2} as 
\begin{align}
\Delta a_{\mu}=(29.0 \pm 9.0)\times 10^{-10},\
\Delta a_{\mu}=(33.5 \pm 8.2)\times 10^{-10}. \label{dev12}
\end{align}
The above results given in Eq. (\ref{dev12}) correspond
to $3.2\sigma$ and $4.1\sigma$ deviations, respectively. 
%%%%
Our formula of muon $g-2$ is given by
\begin{align}
&\Delta a_\mu\approx \frac{m_\mu^2 |f|^2_{22} }{(4\pi)^2}
\biggl[  M F(\psi^{\pm(M+1)},h^{\pm M}) + (M+1)F(h^{\pm M},\psi^{\pm(M+1)})  \biggr],\label{eq:g-2}
\\
%%%
&F(x,y)\approx \frac{
2 m_x^6 +3 m_x^4 m_y^2 - 6 m_x^2 m_y^4 +m_y^6 + 6 m_x^4 m_y^2 \ln\left[\frac{m_y^2}{m_x^2}\right]}
{12(m_x^2- m_y^2)^4}.
\label{damu}
\end{align}
%where we assume $f^a$ is same value for different charged scalar set.
In fig.~\ref{fig:muong-2}, we show the region plot in terms of $f$ and $M_{\psi^{\pm(M+1)}}$ plane for $N=5,7,9,11,13$ cases, where $m_{h^{\pm M}}=380$ GeV and  $N=3$ does not have the allowed region. 
The yellow region satisfies the sizable muon anomalous magnetic moment $2.0\times 10^{-9} \lesssim \Delta a_\mu\lesssim 4.0\times 10^{-9}$. It suggests that the larger number of $N$ provides the larger value of anomalous magnetic moment.

{
Notice here that  the lepton flavor violating (LFV) processes are always arisen in generating the muon anomalous magnetic moment.
The most stringent constraint comes from the $\mu\to e\gamma$ process at the one-loop penguin diagram, and its upper bound of the branching ratio is given by ${\rm BR}(\mu\to e\gamma)\lesssim5.7\times10^{-13}$~\cite{Adam:2013mnn}  at the 95 \% confidential level. The theoretical formulation in our case is computed as
\begin{align}
&{\rm BR}(\mu\to e\gamma)\approx \frac{3\alpha_{\rm em}|(ff^*)_{21}|^2 }{32\pi {\rm G_F^2}}
\left|(
  M F(\psi^{\pm(M+1)},h^{\pm M}) + (M+1)F(h^{\pm M},\psi^{\pm(M+1)})
%g^*g)_{21} F(E,h)-\frac{(f^*f)_{21}}{3 m_{h^\pm}^2}
\right|^2
\label{eq:lfvs}.
\end{align}
To satisfy the constraint, the coupling $(ff^*)_{21}$ has to typically be ${\cal O}(10^{-4})$ under fixed masses; $m_{\Psi^{\pm M}}=100$ GeV, and  $m_{h^{\pm M}}=380$ GeV, where $M$ runs 1 to 6 as we will discuss in the next section.
The result might conflict with the coupling to realize the sizable muon $g-2$ in Fig.~\ref{fig:muong-2}, however we can evade this problem by assuming the off-diagonal  $2-1$ element of $ff^*$ to be zero.
%As can be seen from the above equation, it is generated from the term proportional to $f$, and  its coupling or masses related to exotic fermions or bosons are constrained. 
%%%
%The stringent bound is given by the $\mu\to e\gamma $ process with penguin diagram~\cite{Adam:2013mnn}. However once we can take  $f$ to be diagonal, such LFVs can  simply be evaded.
Even it is the case,  the neutrino mixings are expected to be induced via the coupling of $g\equiv(g_{L/R})$. Hence we retain the consistency of the LFV constraints without conflict of the neutrino oscillation data and the muon anomalous magnetic moment, applying this assumption to the numerical analysis.
}

%\subsection{Muon anomalous magnetic moment}
\section{ Decay processes for the exotic fields}
 Now that we have to consider the decay processes of the exotic fields.
Regardless of the number $M$, $\psi^{-M-1}$ always decays into $\psi^{-M}$ and the charged gauged boson $W^-$. 
And $\psi^{-M}$ decays into $h^{-M}$ and active neutrinos, if the mass of $\psi^{-M}$ is greater than the mass of $h^{-M}$,
or  $h^{-M}$ decays into $\psi^{-M}$ and active neutrinos, if the mass of $\psi^{-M}$ is less than the mass of $h^{-M}$.
Moreover, $k^{-2M}$ can decays into 2$h^{-M}$, if $2m_{h^{\pm M}}< m_{k^{\pm 2M}}$. 
In order to make the analysis simplify, we just assume to be %$m_{h^{-M}} < M_{E^{1/2}}$, and
 $2m_{h^{\pm M}}< m_{k^{\pm 2M}}$.
Therefore all we have to take care of the decay is how to make the $h^{\pm M}$ or $E^{\pm M}$ field decay into the SM fields, which quite depends on the number of $N$. Thus we classify the model in terms of the concrete number of $N$ below.
Notice here that N starts from three, since we assume to be $M\neq0$.

\subsection{N=3}
This is equivalent to $M=1$. 
In this case, we can add to  write the term  
\begin{align}
-{\cal L}_{\rm new} \approx   y_{eE} \bar E_L e_R\varphi + {\rm c.c.},
\end{align}
which suggests the mixing between the SM electric charged leptons and the exotic charged fermions.
The mixing makes the neutrino mass matrix complicate, and $h^\pm$ cannot decay into the SM fields without any additional fields such as another doubly charged boson with $+2$ $U(1)_{\rm B-L}$ charge, and there exist any allowed region to satisfy the muon anomalous magnetic moment in the last section.  Thus we do not mention this case furthermore.

\subsection{N=5}
This is equivalent to $M=2$. 
In this case, we can add to  write the term  
\begin{align}
-{\cal L}_{\rm new} \approx   g' \bar e^c_R e_R  h^{++}  + {\rm c.c.},
\end{align}
that suggest that $h^{--}$ can decay into the same di charged-leptons.
Thus we do not need to add any additional fields,  and decay processes are as follows:
\begin{align}
 h^{--}\to 2\ell^{-}. %\quad E^{--}\to h^{--} +\nu_L.
 \end{align}
Notice here that $g'$ can contribute to the negative contribution of the anomalous magnetic moment, but we can neglect this effect hereafter because this coupling can be take as a  free free parameter.

\subsection{N=7}
This is equivalent to $M=3$. 
In this case, introducing a new field $S^{\pm\pm}$ that is an isospin singlet doubly charged boson with $\pm2$ $U(1)_{\rm B-L}$ charge,
we can add to  write the term  
\begin{align}
-{\cal L}_{\rm new} \approx   g' \bar e^c_R e_R  S^{++}  + y_{eE}  \overline{E_L^{-3}} e_R S^{--} + {\rm c.c.},
\end{align}
where $S^{\pm\pm}$ plays as a role in generating the decaying processes for the exotic fields only.
Then the decay processes are as follows:
\begin{align}
h^{---}\to   E^{---} (+ \nu_L) \to S^{--} (+ \ell^-) \to 2\ell^-.
 \end{align}

\subsection{N=9}
This is equivalent to $M=4$. 
In this case, introducing a new field $S^{\pm\pm}$ that is an isospin singlet doubly charged boson with $\pm2$ $U(1)_{\rm B-L}$ charge,
we can add to  write the term  
\begin{align}
-{\cal L}_{\rm new} \approx   g' \bar e^c_R e_R  S^{++}  + \lambda'  h^{++++} S^{--} S^{--} \varphi + {\rm c.c.},
\end{align}
where $S^{\pm\pm}$ plays as a role in generating the decaying processes for the exotic fields only.
Then the decay processes are as follows:
\begin{align}
h^{----}\to   2 S^{--}  \to 4\ell^-.
 \end{align}

\subsection{N=11}
This is equivalent to $M=5$. 
In this case, introducing new fields ($S^{\pm}$   $S^{\pm\pm}$   $S^{\pm\pm\pm\pm}$)  that are respectively isospin singlet  (singly, double, fourply) charged bosons with the common $\pm2$ $U(1)_{\rm B-L}$ charges,
we can add to  write the term  
\begin{align}
-{\cal L}_{\rm new} &\approx   f' \bar L_L^c L_L S^+ + g' \bar e^c_R e_R  S^{++}   + y_{eE} \overline{E^{-5}_L} e_R S^{----}\nn\\
%%%
&+ \lambda'  S^{++++} + S^{--} S^{--} \varphi^*  + \lambda''  h^{-----} S^{+} S^{++++} \varphi + {\rm c.c.},
\end{align}
where additional fields  play as a role in generating the decaying processes for the exotic fields only.
Then the decay processes are as follows:
\begin{align}
& h^{-----}\to    S^{-}   + S^{----}  \to2 S^{--} (+\nu_L+\ell^-) \to 4\ell^-,\\
& E^{-----}\to       S^{----} (+\ell^-)  \to2 S^{--}  \to 4\ell^-.
 \end{align}

\subsection{N=13}
This is equivalent to $M=6$. 
In this case, introducing new fields ($S^{\pm\pm}$   $S^{\pm\pm\pm\pm}$)  that are respectively isospin singlet  (double, fourply) charged bosons with the common $\pm2$ $U(1)_{\rm B-L}$ charges,
we can add to  write the term  
\begin{align}
-{\cal L}_{\rm new} &\approx  g' \bar e^c_R e_R  S^{++}  
+ \lambda'  S^{++++} + S^{--} S^{--} \varphi^*  + \lambda''  h^{------} S^{++} S^{++++} \varphi + {\rm c.c.},
\end{align}
where additional fields  play as a role in generating the decaying processes for the exotic fields only.
Then the decay processes are as follows:
\begin{align}
& h^{------}\to    S^{--}   + S^{----}  \to6 \ell^-.
 \end{align}

\section{Diphoton excess}
%%%%%%%%%%%%%%%%%%%
\begin{figure}[tb]
\begin{center}
\includegraphics[width=60mm]{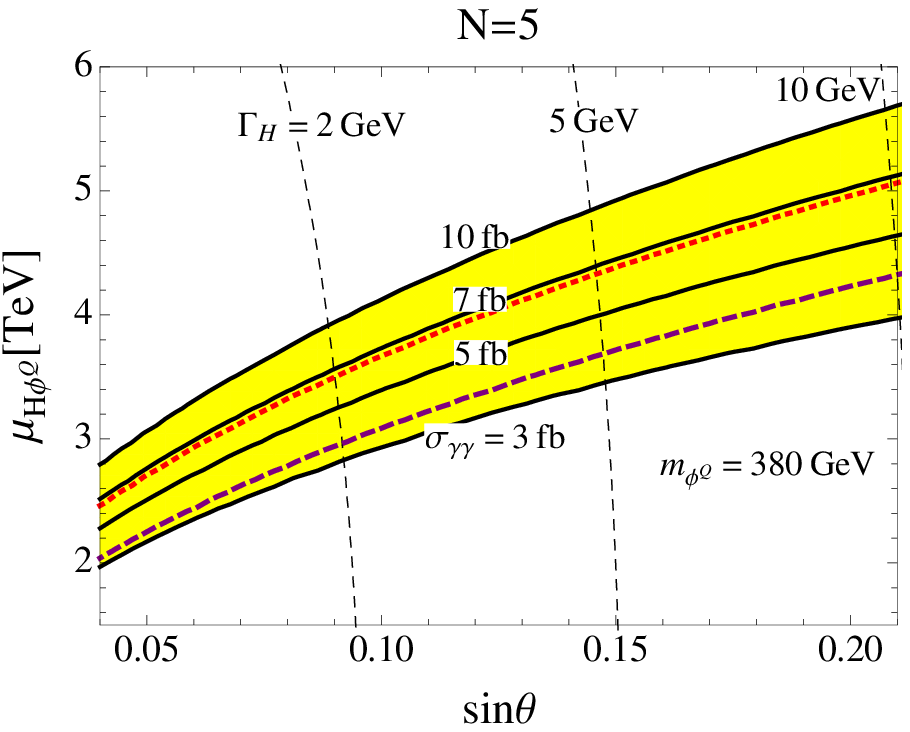} \qquad
\includegraphics[width=60mm]{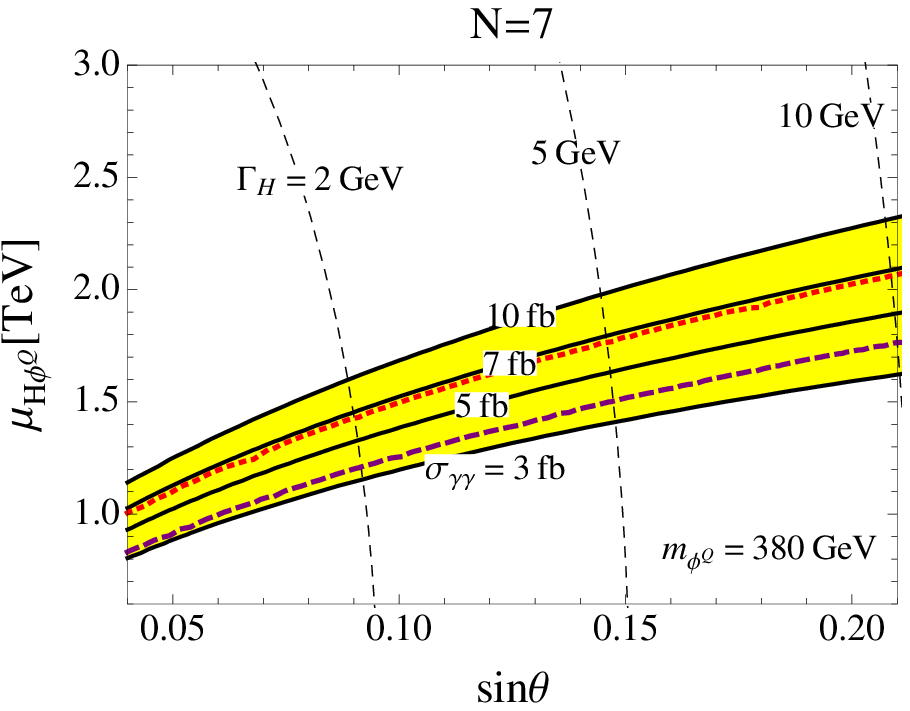}
\includegraphics[width=60mm]{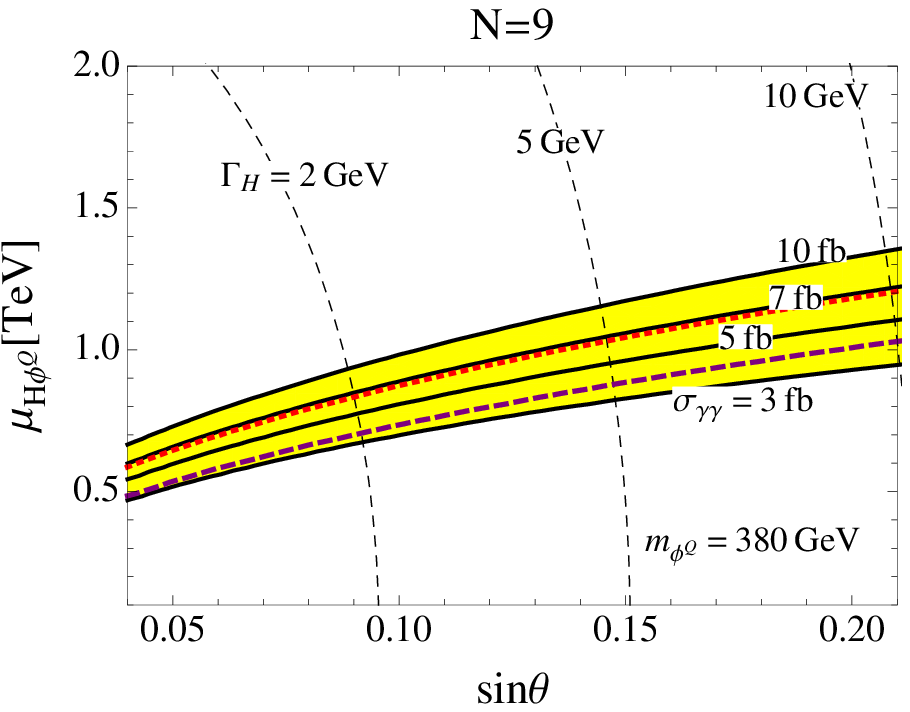} \qquad
\includegraphics[width=60mm]{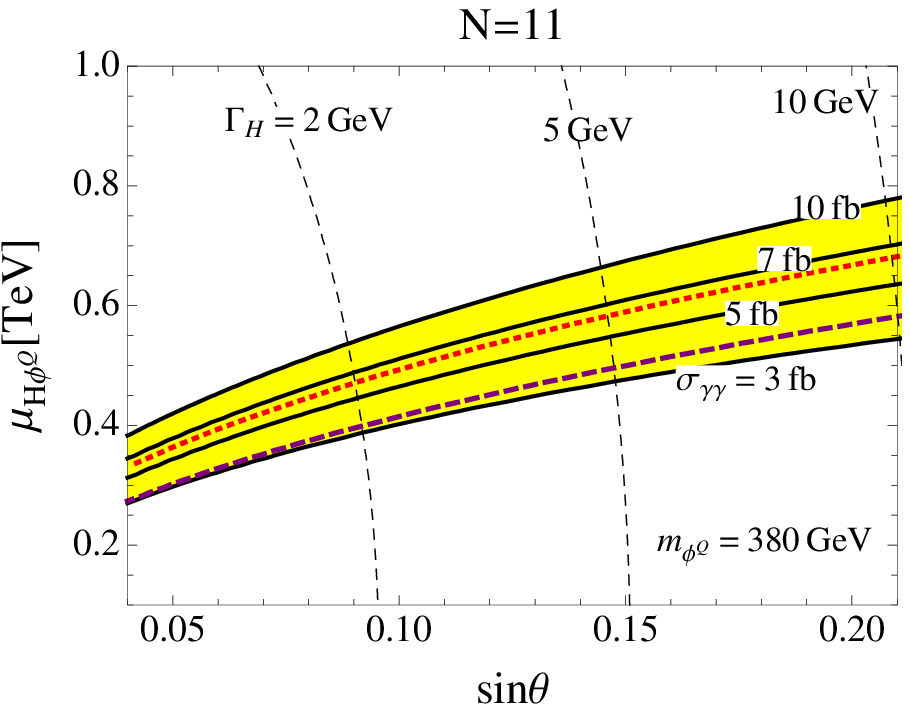}
\includegraphics[width=60mm]{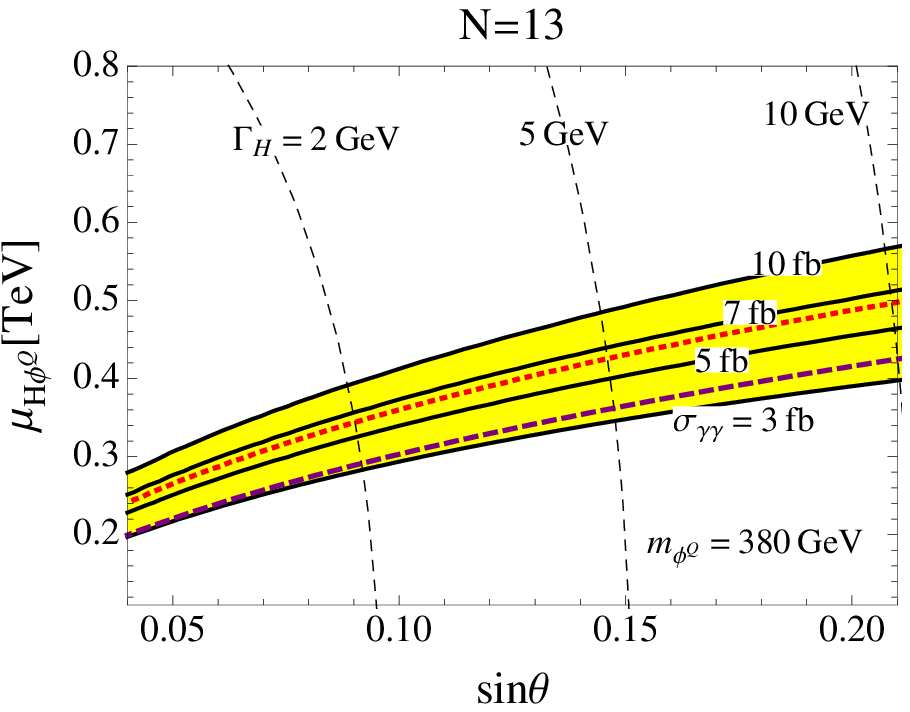}
\caption{The contours of $\sigma (gg \to H) BR(H \to \gamma \gamma)$ (in unit of fb) and total width $\Gamma_H$ (in unit of GeV) in $\sin \theta - \mu_{H \phi^Q}$ plane for $N= 5,7,9,11,13$.
All the charged scalar masses are taken to be 380 GeV. The purple dashed and the red dotted lines indicate constraint from diphoton search at 8 TeV for $R_{\gamma \gamma} =2$ and $4$ where the region above the lines are excluded.
}   \label{fig:diphoton1}
\end{center}\end{figure}
%%%%%%%%%%%%%%%%%%%
We discuss the diphoton excess in case of $N=5,7,9,11,13$ as discussed in the previous section where the candidate of 750 GeV scalar boson is heavy CP even scalar $H$.
The couplings relevant to diphoton decay are obtained from quartic couplings including $\Phi$, $\varphi$ and charged scalar fields;
\begin{equation}
{\cal L} \supset \sum_{i} \left[ \lambda_{\Phi \phi_i^Q} |\Phi|^2 |\phi_i^Q|^2 + \lambda_{\varphi \phi_i^Q} | \varphi|^2 |\phi^Q_i|^2 \right], 
\end{equation}
where we denote charged scalar field with electric charge $Q$ as $\phi^Q_i$ in general. 
After $\Phi$ and $\varphi$ get VEV, relevant interactions for mass eigenstates are 
\begin{equation}
{\cal L} \supset  \sum_i  \left[ (\lambda_{\Phi \phi_i^Q} v \cos \theta - \lambda_{\varphi \phi_i^Q} v_\phi \sin \theta) h |\phi^Q_i|^2 + (\lambda_{\Phi \phi_i^Q} v \sin \theta + \lambda_{\varphi \phi_i^Q} v_\phi \cos \theta) H |\phi^Q_i|^2 \right].
\end{equation}
Since we want to suppress contribution to $h \to \gamma \gamma$, we assume $\lambda_{\Phi \phi^Q_i} v \cos \theta \simeq \lambda_{\varphi \phi^Q_i} v_\phi \sin \theta$.
Then interactions contributing to $H \to \gamma \gamma $ become
\begin{equation}
\sum_i \frac{\lambda_{\varphi \phi^Q_i} v_\phi}{\cos \theta} H |\phi^Q_i|^2 = \sum_i \mu_{H \phi^Q_i} H |\phi^Q_i|^2,
\end{equation}
where $\mu_{H \phi^Q_i} = \lambda_{\varphi \phi^Q_i} v_\phi/ \cos \theta$.

The heavy CP-even scalar $H$ can be produced by gluon fusion process via mixing with SM Higgs. The cross section is then obtained as~\cite{Djouadi:2013uqa,Khachatryan:2014wca}
\begin{equation}
\sigma(gg \to H) \simeq \sin^2 \theta \times 0.85 \ {\rm pb},
\end{equation}
at the LHC 13 TeV.
Furthermore, $H$ can be produced by photon fusion, $pp(\gamma \gamma) \to H$, since effective $H \gamma \gamma$ coupling would be sizable due to charged scalar loop contributions.
Here we apply the estimation of the cross section for photon fusion including both elastic and in-elastic scattering in Ref.~\cite{Csaki:2016raa}
\begin{equation}
\label{eq:photon-fusion}
\sigma(pp (\gamma \gamma) \to H \to \gamma \gamma +X)_{\rm 13 TeV} = 10.8{\rm pb} \left( \frac{\Gamma_H}{45 {\rm GeV}}  \right) \times BR^2(H \to \gamma \gamma),
\end{equation}
where $X$ indicate any other associated final states. 
Therefore total cross section for $pp \to H \to \gamma \gamma$ is obtained as
\begin{equation}
\sigma_{\gamma \gamma} = \sigma (gg \to H) BR(H \to \gamma \gamma) + \sigma_{\gamma-{\rm fusion}}
\end{equation}
where $\sigma_{\gamma-{\rm fusion}}$ is given by Eq.~(\ref{eq:photon-fusion}).

Through mixing with SM Higgs, $H$ decays into SM particles where the dominant partial decay widths are:
\begin{align}
\Gamma(H \to W^+ W^-) =&  \frac{g^2 m_W^2 \sin^2 \theta}{64 \pi m_{H}} \frac{m_{H}^4 - 4 m_{H}^2 m_W^2 + 12 m_W^2}{m_W^4} \sqrt{1- \frac{(2 m_W)^2}{m_{H}^2}}\,,  \\
\Gamma(H \to Z Z) =& \frac{1}{2} \frac{g^2 m_Z^2 \sin^2 \theta}{64 \pi \cos^2 \theta_W m_{H}} \frac{m_{H}^4 - 4 m_{H}^2 m_Z^2 + 12 m_Z^2}{m_Z^4} \sqrt{1- \frac{(2 m_Z)^2}{m^2_{H}} }\,,  \\
\Gamma(H \to t \bar t) = & \frac{3 m_t^2 \sin^2 \theta}{8 \pi v^2} m_\phi \sqrt{1 - \frac{4 m_{t}^2}{m_H^2}}.
\end{align}
Here partial decay widths for other fermion channels are subdominant.
The decay process $H \to \gamma \gamma $ is induced by charged particle loops.
The partial decay width is given by
\begin{equation}
\Gamma_{H \to \gamma \gamma} \simeq \frac{\alpha^2 m_\phi^3}{256 \pi^3} \left| \sum_{\phi^{Q}_i} Q_i^2  \frac{\mu_{H \phi^Q_i} }{2 m_{\phi^Q_i}^2} A_0 (\tau_{\phi^Q_i}) \right|^2,
\end{equation}
where $A_0 (x) = -x^2[x^{-1} - [\sin^{-1} (1/\sqrt{x})]^2]$ and $\tau_{\phi^Q_i} = 4 m_{\phi^Q_i}^2/m_H^2$ and we omit SM particle contribution since they are small compared with charged scalar contributions. 
Note that we also have $H \to Z\gamma $ mode which is subdominant contribution and we omit the formula for decay width here.

The constraint from 8 TeV data should be taken into account.
The most stringent constraint in our scenario is given by diphoton search at 8 TeV since our $BR(H \to \gamma \gamma)$ is large.
The constraint is~\cite{CMS:2014onr,  Aad:2015mna}
\begin{equation}
\sigma_{\gamma \gamma}^{\rm 8 TeV} \equiv \sigma (gg \to H)^{\rm 8 TeV} BR(H \to \gamma \gamma) + \sigma_{\gamma-{\rm fusion}}^{\rm 8 TeV} < 1.5 \ {\rm fb}.
\end{equation}
Here the ratios of 13 TeV cross section and that of 8 TeV are written as $\sigma(gg \to H)^{\rm 13 TeV}/\sigma(gg \to H)^{\rm 8 TeV} \simeq 5$~\cite{Franceschini:2015kwy} and $\sigma_{\gamma-{\rm fusion}}^{\rm 13 TeV}/\sigma_{\gamma-{\rm fusion}}^{\rm 8 TeV} \equiv R_{\gamma \gamma}$. The $R_{\gamma \gamma}$ is estimated to be $\sim 2$ but the uncertainty is large so that it can be $\sim 4$~\cite{Fichet:2015vvy,Csaki:2016raa}. Thus we investigate the constraint with $R_{\gamma \gamma}=2$ and 4.

We then estimate the product of $H$ production cross section and branching ratio for diphoton channel $\sigma_{\gamma \gamma}^{\rm total}$ for the cases of $N=3,5,7,9$ and $11$.
For simplicity, we assume couplings $\mu_{H \phi^Q_i}$ take same value for all charged scalars. 
Also we choose mass of charged scalar as $m_{\phi^Q_i} = 380$ GeV to enhance loop function inside the diphoton decay width.
The contours of $\sigma_{\gamma \gamma}^{\rm total}$ and $\Gamma_H$ are shown in Fig.~\ref{fig:diphoton1} by solid and dashed lines respectively for $N= \{ 5,7, 9,11,13 \}$, 
and the purple dashed and the red dotted lines indicate constraint from diphoton search at 8 TeV for $R_{\gamma \gamma} =2$ and $4$ where the region above the lines are excluded.
We find that small $\sin \theta$ region is strongly constrained for $R_{\gamma \gamma }=2$ but we can obtain $\sim 3$ fb cross section for all $\sin \theta$. 
On the other hand, we can obtain $\sim 5$ fb cross section for $R_{\gamma \gamma}=4$.
Furthermore the trilinear coupling can be less than $1$ TeV for $7\le N$. 
We also find that the total decay width $\Gamma_H$ is $O (1 - 10)$ GeV for the parameter region which explain the diphoton excess.

%\section{Conclusions}
\section{ Conclusions and discussions}
In our paper, we have proposed a generalized Zee-Babu model with a global  $U(1)$ B-L symmetry, in which neutrino masses are induced at the two loop level. Here we have introduced vector like isospin doublet fermions with general $N/2$ hypercharges that can explain the discrepancy of the muon anomalous magnetic moment to the standard model (SM) sizably.  Corresponding to each of the number of $N$, we have needed to introduce some multiply charged bosons in order to make the exotic fields decay into the standard model fields.
Diphoton excess is explained by introducing several charged bosons, depending on the number of $N$.
We have classified the model according to the number of $N$, and we have discussed what kind of models  are in favor of explaining the sizable diphoton resonance as well as the muon anomalous magnetic moment.
 
We have estimated the product of $H$ production cross section and branching ratio for diphoton channel.
Then the $O(1)$ TeV trilinear coupling is required for $N=5$ while the coupling can be smaller for larger $N$.
Thus larger $N$ is preferred to satisfy tree level unitarity and explain the diphoton excess.
We also find the total decay width in our scenario is $O(1- 10)$ GeV.
Furthermore we investigated constraint from diphoton search at 8 TeV and we find the parameter region which explain diphoton excess and can satisfy the constraint.

A Dirac type of dark matter can be involved in our theory without conflict of any phenomenological point of views~\cite{Lindner:2011it}.
However this analysis is beyond the scope.

%However since the dark matter does not link to neutrinos, muon anomalous magnetic moment, and the diphoton excess, we just refer to the ref.~\cite{Lindner:2011it} instead of the discussion in details.

%\newpage
%%%%%%%%%%%%%%%%%%%%%%%%%%%%%%%%%%%
%\hspace{0.2cm} {\bf Acknowledgments}
%\section*{Acknowledgments}:
%\vspace{0.5cm}
\section*{Acknowledgments}
\vspace{0.5cm}
H.O. thanks to Prof. Shinya Kanemura, Prof. Seong Chan Park, Dr. Kenji Nishiwaki, Dr. Yuta Orikasa, Dr. Ryoutaro Watanabe, and Dr. Kei Yagyu for fruitful discussions. 
H. O. is sincerely grateful for all the KIAS members, Korean cordial persons, foods, culture, weather, and all the other things.
%%%%%%%%%%%%%%%%%%%%%%%%%%%%%%%%%%%
%%%%%%%%%%%%%%%%%%%%%%%%%%%%%%%%%%%

\end{document}